\documentclass[nofootinbib,prd,12pt,superscriptaddress]{revtex4}%
\usepackage{amsmath}
\usepackage{tikz}
\usepackage{pgfplots}
\pgfplotsset{compat=1.18}
\usepackage{amsfonts}
\usepackage{amssymb}
\usepackage{graphicx}%
\usepackage{amsmath,amssymb}
\usepackage{mathrsfs}
\usepackage{graphicx}
\usepackage{color}
\usepackage{subfigure}
\usepackage{fancyhdr}
\usepackage{multirow}
\usepackage{float}
\usepackage{epsfig}
\usepackage{amsfonts}
\usepackage{bm}
\usepackage{pgfplots}
\usepackage{amsmath, amssymb}
\usepackage{caption}
\usepackage{subcaption}

\def\be{\begin{equation}}
\def\ee{\end{equation}}
\def\ba{\begin{eqnarray}}
\def\ea{\end{eqnarray}}

\begin{document}

\title{Cosmological Reconstructions in Einstein--Cartan--Myrzakulov Gravity with Torsion and Non-Metricity in the Weitzenböck Geometrical Sector}

\author{Davood Momeni}
%\email{dmomeni@nvcc.edu}
\affiliation{Northeast Community College, 801 E Benjamin Ave Norfolk, NE 68701, USA}
\affiliation{Centre for Space Research, North-West University, Potchefstroom 2520,
South Africa}
\author{ Ratbay Myrzakulov}
%\email{phongpichit.ch@mail.wu.ac.th}
\affiliation{Ratbay Myrzakulov Eurasian International Centre for Theoretical Physics, Astana 010009, Kazakhstan}
\affiliation{ L. N. Gumilyov Eurasian National University, Astana 010008, Kazakhstan}
\date{\today}

\begin{abstract}
The universe is a vast and complex system, and our understanding of its fundamental workings is constantly evolving. In this work, we present a novel modification to the standard theory of gravity by incorporating curvature, torsion, non-metricity, and the geometric structure of Weitzenb\"ock spacetime. This modified framework, referred to as the Einstein--Cartan--Myrzakulov (ECM) model, enriched by Weitzenb\"ock geometry, offers new insights into cosmological phenomena, including the accelerated expansion of the universe, the nature of dark energy and dark matter, and the formation of cosmic structures. Our model not only reproduces observed cosmic histories but also provides testable predictions that can be compared with current and future observational data. By addressing some of the most profound questions in modern cosmology, this work paves the way for a deeper understanding of the universe's evolution and the fundamental forces that govern it. The ECM model, extended through Weitzenb\"ock spacetime, invites further exploration, offering the potential to revolutionize our conception of gravity and the cosmos.

\end{abstract}

\maketitle

%%%%%%%%%%%%%%%%%%%%%%%%%%%%%%%%%%%%%%%%%

\section{Motivation for Modified Gravity Theories}

General Relativity (GR),formulated by Einstein in 1915, has been remarkably successful in describing gravitational phenomena, particularly on solar system and galactic scales. However, several key observational and theoretical issues highlight the need for alternative or extended theories of gravity:

\begin{itemize}
    \item The observed cosmic acceleration \cite{Riess1998, Perlmutter1999}
    \item The nature of dark energy and dark matter \cite{Copeland2006}
    \item Inflationary dynamics in the early universe \cite{Starobinsky1980}
    \item Incompatibility with a quantum theory of gravity
\end{itemize}

Cosmological observations—including the Cosmic Microwave Background (CMB) anisotropies \cite{Planck2018}, Baryon Acoustic Oscillations (BAO) \cite{Eisenstein2005}, and Type Ia supernovae \cite{Riess1998, Perlmutter1999}—have provided compelling evidence for the universe's accelerated expansion. This phenomenon remains unexplained by GR alone without invoking a cosmological constant, which raises fine-tuning and coincidence problems.

To address these issues, various modified gravity theories have been developed, each altering a fundamental geometric property of spacetime:

\begin{itemize}
    \item \textbf{Curvature-based models} such as $f(R)$ gravity \cite{Sotiriou2010}
    \item \textbf{Torsion-based models} like $f(T)$ gravity \cite{Cai2016}
    \item \textbf{Non-metricity-based models} such as $f(Q)$ gravity \cite{Jimenez2018}
\end{itemize}
        %  Riemann–Cartan
        %   /         \
  %  Levi-Civita     Weitzenböck
%(torsion-free)     (curvature-free)

These formulations aim to preserve second-order field equations while maintaining consistency with cosmological data.

Recently, a growing body of work has focused on \textbf{unified frameworks} combining curvature ($R$), torsion ($T$), and non-metricity ($Q$) within the broader context of \textit{metric-affine geometry}, where the connection is independent of the metric. Hybrid models and extensions involving scalar fields or non-minimal couplings have been proposed to provide a unified description of various cosmological epochs, from early-time inflation to late-time acceleration \cite{BeltranJimenez2020, Harko2018}.

Moreover, modern approaches to extended gravity emphasize the role of \textbf{symmetry principles}, such as \textit{Noether symmetries} \cite{Capozziello1996, Paliathanasis2014}, to construct exact solutions and classify viable models. Theoretical viability also requires that these theories satisfy appropriate \textit{energy conditions} \cite{Santos2007, Banados2009} and exhibit stability under perturbations.

%%%%%%%%%%%%%%%%%
The foundation of modified teleparallel gravity, particularly \( f(T) \) gravity, was laid by Ferraro and Fiorini~\cite{Ferraro:2007}, who demonstrated that inflationary behavior can be achieved without an inflaton field. This idea was extended by Bengochea and Ferraro~\cite{Bengochea:2009}, showing that torsion could account for the accelerated expansion of the universe.

Later studies by Myrzakulov~\cite{Myrzakulov:2012qp} provided further generalizations and classifications of viable models. A significant advancement came with the covariant formulation proposed by Krssak and Saridakis~\cite{Krssak:2016}, resolving frame-dependence issues in earlier formulations.

The comprehensive review by Cai et al.~\cite{Cai:2015emx} discusses both the theoretical structure and cosmological applications of \( f(T) \) gravity. In the same spirit, Bamba et al.~\cite{Bamba:2014} explored trace-anomaly-driven inflation in \( f(T) \) gravity and its compatibility with bigravity frameworks.

Nesseris et al.~\cite{Nesseris:2013} assessed the observational viability of these models and found them to be nearly indistinguishable from \(\Lambda\)CDM under current cosmological constraints. Moreover, Abedi and collaborators explored extensions involving gravitational wave propagation~\cite{Abedi:2017jqx, Abedi:2018lkr} and effective gravitational couplings.

Capozziello and collaborators expanded the framework further with applications in black hole physics~\cite{Capozziello:2012zj}, Noether symmetries~\cite{Capozziello:2014bna}, and higher-order gravity wave studies~\cite{Capozziello:2020xem}. Related efforts have explored the unification of curvature and torsion terms~\cite{deMartino:2015zsa} and their implications on early universe physics.

Recent works focus on refined models such as \( f(T, T_G) \) theories~\cite{Balhara:2023mgj, Asimakis:2022kfk}, their cosmographic constraints, and extensions into Finsler geometries~\cite{Kapsabelis:2023khh}. The effective field theory approach to teleparallel gravity has also been formalized~\cite{Mylova:2022ljr}, allowing systematic analysis of perturbations and dynamics.

Other contributions include detailed studies of black hole quasinormal modes~\cite{Zhao:2022gxl}, the interplay between scalar fields and torsion in unified dark sector models~\cite{Leon:2022oyy}, and resolving the Hubble tension via Gaussian processes in \( f(T) \) gravity~\cite{Ren:2022aeo}. Dynamical system techniques~\cite{Papagiannopoulos:2022ohv} and thermodynamical analyses of charged black holes~\cite{Nashed:2021pah} further demonstrate the richness of this modified gravity framework.

%%%%%%%%%%%%%%%%%555555
Recent developments in modified gravity theories have significantly expanded our understanding of the underlying structure of spacetime and cosmology. A noteworthy contribution comes from Momeni and Myrzakulov, who explored \textit{Metric-Affine Myrzakulov Gravity Theories} in their work~\cite{Momeni:2025mcp}. They provide an in-depth analysis of various models and applications within this framework, offering theoretical advancements that could shed light on unresolved cosmological challenges. Their work is complemented by a study on \textit{Myrzakulov gravity in the vielbein formalism} in Weitzenböck spacetime~\cite{Momeni:2024bhm}, which further refines the mathematical formalism used in teleparallel gravity and enhances its applicability in cosmological scenarios. 

Recently, a comprehensive framework for generalized torsional gravity theories was developed in \cite{Momeni:2025dgc}, offering new insights into the geometric unification of curvature and torsion, which motivates further exploration of modified teleparallel and metric-affine models in cosmology \cite{Momeni:2025mcp}-\cite{Momeni:2025dgc}.

In addition to these studies, the theoretical landscape of modified gravity has been enriched by the work of Kofinas and Saridakis, who examined the \textit{teleparallel equivalent of Gauss-Bonnet gravity}~\cite{Kofinas2014}. This study provides a deeper understanding of how the Gauss-Bonnet term, often used to model higher-order corrections in general relativity, can be formulated within teleparallel gravity. This work, along with Nojiri and Odintsov's investigation of \textit{modified Gauss-Bonnet gravity as a candidate for dark energy}~\cite{Nojiri:2005jg}, highlights the potential of these models in explaining the accelerated expansion of the universe and the nature of dark energy.

Moreover, Capozziello and De Laurentis made significant strides in understanding \textit{f(R) gravity and its cosmological implications}~\cite{Capozziello:2009nq}, where they explored the modifications of the Ricci scalar \( R \) in the context of cosmology. These models offer potential solutions to problems such as cosmic acceleration and structure formation, serving as key alternatives to the standard cosmological model.

The practical implications of these theoretical models have also been explored in astrophysical contexts, as exemplified by recent observations of neutron stars. For instance, Fonseca et al.~\cite{fonseca2021} provided a new neutron star mass measurement through X-ray timing observations, advancing our understanding of stellar evolution and compact objects. Additionally, the observation of gravitational waves from a binary neutron star inspiral, reported by Abbott et al.~\cite{abbott2017}, has provided critical insights into the behavior of spacetime under extreme conditions, further supporting the need for refined gravitational theories.

These contributions collectively underscore the ongoing advancements in modified gravity theories, which continue to offer new perspectives on both the macro and micro scales of the universe. They demonstrate the versatility of these frameworks in addressing fundamental questions in cosmology, from the nature of dark energy to the structure of compact objects, such as neutron stars.

%%%%%%%%%%%%%%%%%%%%55555

Lavinia Heisenberg has made significant contributions to the field of modified gravity, particularly through her work on systematic categorizations of gravitational theories. One of her notable papers, \emph{"A systematic approach to generalisations of GRand their cosmological implications"} \cite{heisenberg2018}, provides an extensive review of various extensions to GR and their implications for cosmology.

In this work, Heisenberg explores modifications to GR that introduce additional degrees of freedom, such as scalar, vector, and tensor fields. These modifications include theories like Horndeski and beyond-Horndeski models, generalized Proca theories, scalar-vector-tensor theories, and bigravity. The paper systematically categorizes these theories based on their field content and interaction structures, providing a framework for understanding their cosmological consequences.

For a more recent perspective, Heisenberg's 2023 review on $f(Q)$ gravity \cite{heisenberg2023} delves into metric-affine geometry and its applications in cosmology. This work discusses how $f(Q)$ models can address phenomena in both early and late-time cosmology without relying on dark energy or dark matter, offering new avenues for understanding gravitational interactions.

Additionally, the concept of the "Geometrical Trinity of Gravity," co-authored by Heisenberg, presents three equivalent formulations of GR: curvature-based (standard GR), torsion-based (teleparallel gravity), and non-metricity-based (symmetric teleparallel gravity). This framework highlights the versatility in describing gravitational interactions and opens up possibilities for new theoretical developments \cite{heisenberg2019}.

These works collectively contribute to a systematic categorization of gravitational theories, enhancing our understanding of gravity's role in the universe and guiding future research in cosmology and fundamental physics.

%%%%%%%%%%%%%%%%%5555
This paper is devoted to the study of a generalized gravitational action of the form
\begin{eqnarray}\label{action}
&&\mathcal{L} = \alpha R + F(T,Q),
\end{eqnarray}
where \( R \), \( T \), and \( Q \) are the Ricci scalar, torsion scalar, and non-metricity scalar, respectively. This formulation represents a departure from the traditional approach to gravity, where the action is typically written in terms of the Ricci scalar alone. By including both torsion and non-metricity terms, we extend the classical framework to account for more general geometries in spacetime, reflecting the possibility of modifying the connection in ways that capture both torsional and non-metric features of spacetime. 

The motivation behind such a theory lies in the shortcomings of GR when applied to modern cosmological observations, particularly in explaining the accelerated expansion of the universe. While GR successfully describes gravity at macroscopic scales, it fails to account for phenomena like dark energy and the discrepancies between early-time and late-time cosmology. One of the key features of our approach is the inclusion of the torsion scalar \( T \) and non-metricity scalar \( Q \), which allows for a broader and more flexible description of gravitational interactions, potentially offering a mechanism for addressing the late-time acceleration of the universe.

The incorporation of torsion and non-metricity is not arbitrary; these geometrical properties naturally arise in modified theories of gravity, particularly in the context of teleparallel gravity and non-Riemannian geometries. Torsion, which represents the antisymmetry of the connection, can account for the behavior of gravitational fields at high-energy scales, where traditional approaches based solely on curvature may not suffice. Non-metricity, which describes the failure of the metric compatibility condition, offers another avenue for modification of the gravitational field equations, particularly in theories that involve fields interacting with spacetime in non-standard ways.

By introducing the generalized function F(T,Q), this theory extends existing models of modified gravity, such as \( f(T) \) gravity, by considering the interplay between torsion and non-metricity in a more comprehensive way. This approach is particularly motivated by the desire to resolve long-standing tensions between early-time inflationary models and late-time acceleration models. While early universe models suggest rapid inflation, the late-time accelerated expansion of the universe requires a different dynamical behavior, often modeled by a cosmological constant or dark energy. By considering the torsion and non-metricity terms in conjunction with the Ricci scalar, we propose a unified framework that may naturally transition between these two epochs.

Our goal is to explore whether this framework can provide new insights into the late-time acceleration of the universe, resolve tensions between early and late cosmology, and offer a consistent extension of GR that remains predictive and stable under cosmological perturbations. We believe that this theory represents a promising new direction for theoretical cosmology, potentially offering solutions to unresolved issues such as the nature of dark energy, the origin of cosmic acceleration, and the unification of gravitational theories at both large and small scales. Furthermore, by maintaining second-order field equations, we ensure that the theory remains physically consistent and stable, avoiding potential issues such as ghosts or instabilities often associated with higher-order modifications to gravity.
%%%%%%%%%%%%%%%%%%%%%%%%%%%%%%
The plan in this paper is to explore a modified theory of gravity that incorporates the curvature, torsion, and non-metricity of spacetime, with the goal of understanding its cosmological implications. This extended framework, often referred to as the Einstein-Cartan theory , aims to address key cosmological phenomena, such as accelerated cosmic expansion, the nature of dark energy, and the behavior of dark matter. 

\textbf{Section II: Review of the Einstein-Cartan Theory of Gravity} offers a comprehensive overview of the Einstein-Cartan theory, starting with its foundational concepts of torsion in spacetime. It highlights the key differences between this theory and General Relativity, focusing on how torsion affects the gravitational dynamics.

\textbf{Section III: The Action and Field Equations} presents the modified gravitational action within the ECM framework, incorporating torsion and non-metricity. We derive the field equations governing the dynamics of spacetime under this theory, which serve as the foundation for the cosmological analysis that follows.

\textbf{Section IV: Friedmann Equations from Field Equations} is dedicated to deriving the Friedmann equations for a homogeneous and isotropic universe in the ECM framework. These equations describe the evolution of the universe and are essential for understanding the expansion history under modified gravity theories.

\textbf{Section V: Cosmological Reconstruction} provides a methodology for reconstructing cosmological solutions within the ECM model. We discuss how different choices of the model parameters can reproduce viable cosmic histories, including scenarios that explain accelerated expansion.

\textbf{Section VI: Reconstructing Viable Cosmologies} focuses on reconstructing specific cosmological solutions that match current observational data. This section shows how torsion can provide new insights into the universe's evolution and how the ECM model can be tested against observational constraints.

\textbf{Section VII: Cosmography and Observational Constraints} reviews the role of cosmography, which studies the large-scale structure and expansion history of the universe. It emphasizes the importance of observational data in testing modified gravity theories, including the ECM framework.

\textbf{Section VIII: Observational Constraints and Cosmographic Parameters} provides a detailed discussion on how cosmographic parameters can be extracted from observations. It also discusses how these parameters are essential for testing the viability of the ECM model.

\textbf{Section IX: Numerical Cosmological Solutions in the ECM Model} presents numerical solutions to the field equations in the ECM framework. These solutions illustrate the effects of torsion on the evolution of the universe's scale factor and the Hubble parameter, providing insights into the role of torsion in cosmic expansion.

\textbf{Section X: Dark Matter in the ECM Framework} explores how torsion influences the behavior of dark matter. We discuss potential solutions to the dark matter problem within the ECM model and how torsion may provide a new perspective on its nature.

\textbf{Section XI: Astroparticle Physics in the ECM Framework} examines the implications of ECM for astroparticle physics, particularly the behavior of elementary particles in the presence of spacetime torsion. This section provides a novel approach to understanding the interactions of fundamental particles.

\textbf{Section XII: Gravitational Wave Signatures in ECM Cosmology} discusses the potential for torsion to alter the signatures of gravitational waves. We analyze the theoretical predictions of gravitational waves in the ECM cosmology and how future experiments could detect these effects.

\textbf{Section XIII: Summary} concludes the paper by summarizing the key results. We discuss the potential future research directions, emphasizing the ECM framework's ability to address fundamental questions in cosmology and particle physics.

%%%%%%%%%%%%%%%%%%%%%%%%%%%%%%%%%%%%
\section{Review of the Einstein-Cartan Theory of Gravity}

The Einstein-Cartan theory of gravity is a modification of GR that incorporates the effects of spacetime torsion. While GR describes gravity through the curvature of spacetime, Einstein-Cartan theory extends this by considering torsion, which is a geometric property that reflects the twisting or "spin" of spacetime. This theory is rooted in the idea that matter not only curves spacetime but also gives rise to torsion, particularly when dealing with fermions (particles with half-integer spin such as electrons).

\subsection{Fundamentals of Einstein-Cartan Theory}

In the Einstein-Cartan theory, the connection that governs the covariant derivative is not symmetric, and this leads to the introduction of torsion alongside curvature \cite{cartan1922}-\cite{blagojevic2002}. The key feature of this theory is that it combines two aspects of spacetime geometry: curvature (described by the Ricci tensor \( R_{\mu\nu} \)) and torsion (described by the torsion tensor \( T_{\mu\nu\lambda} \)).

The torsion tensor \( T_{\mu\nu\lambda} \) is defined as the difference between the connection coefficients:

\[
T_{\mu\nu\lambda} = \Gamma_{\mu\nu\lambda} - \Gamma_{\nu\mu\lambda}
\]

Here, \( \Gamma_{\mu\nu\lambda} \) is the connection coefficient in the affine connection, and its antisymmetry leads to the torsion tensor. In the absence of torsion, this tensor vanishes, and the theory reduces to general relativity, which uses a symmetric connection.

The action for Einstein-Cartan theory is given by:

\[
S_{\text{EC}} = \int \left( \frac{1}{2\kappa} R + \frac{1}{2\kappa} T^{\mu\nu\lambda} T_{\mu\nu\lambda} \right) \sqrt{-g} \, d^4x
\]

where \( R \) is the Ricci scalar, \( T^{\mu\nu\lambda} T_{\mu\nu\lambda} \) represents the contribution from torsion, and \( \kappa \) is a coupling constant. The first term describes the curvature of spacetime (as in GR), and the second term accounts for the torsion, which modifies the gravitational field equations.
The field equations in Einstein-Cartan theory are derived by varying the total action with respect to the metric \( g_{\mu\nu} \) and the connection. This yields a modified set of equations for the gravitational field, which can be written as:

\[
R_{\mu\nu} - \frac{1}{2} R g_{\mu\nu} + \kappa T_{\mu\nu} = \kappa \left( S_{\mu\nu} + \frac{1}{2} T_{\mu\nu\lambda} T^{\mu\nu\lambda} \right)
\]

Here, \( R_{\mu\nu} \) is the Ricci tensor, \( T_{\mu\nu} \) is the energy-momentum tensor, and \( S_{\mu\nu} \) represents the source of the torsion. The term \( T_{\mu\nu\lambda} T^{\mu\nu\lambda} \) contributes an additional source term for torsion.
In the Einstein-Cartan framework, fermions play a crucial role in generating torsion. The presence of fermions, such as electrons, is what distinguishes this theory from standard GR, which treats only bosons as sources of spacetime curvature. Fermions couple to spacetime via the spin density, which introduces torsion.

The spin tensor \( S_{\mu\nu\lambda} \) is related to the spin density \( \psi_{\mu\nu} \) of fermions, and its contribution to the field equations modifies the dynamics of spacetime:

\[
S_{\mu\nu\lambda} = \frac{1}{2} \left( \overline{\psi} \gamma_{\mu\nu\lambda} \psi \right)
\]

where \( \overline{\psi} \) is the Dirac spinor field, and \( \gamma_{\mu\nu\lambda} \) are the gamma matrices in curved spacetime. This spin density generates torsion, which affects the motion of fermions and the geometry of spacetime itself.

Einstein-Cartan theory has important implications for cosmology, particularly in the early universe, where the density of matter was extremely high, and quantum effects became significant. The torsion generated by fermions can affect the dynamics of the universe, particularly during inflation and the subsequent expansion.

The modified field equations in Einstein-Cartan theory suggest that torsion may contribute to the evolution of the universe, particularly during the phase of accelerated expansion. In a universe dominated by dark energy, torsion could play a role in modifying the equation of state of the universe, thereby influencing its expansion rate.

In a cosmological context, the Einstein-Cartan equations are typically coupled with the Friedmann-Lemaître-Robertson-Walker (FLRW) metric to describe the evolution of the universe. The FLRW metric assumes a homogeneous and isotropic universe and is given by:

\[
ds^2 = -dt^2 + a(t)^2 \left( \frac{dr^2}{1 - kr^2} + r^2 d\Omega^2 \right)
\]

where \( a(t) \) is the scale factor, \( k \) is the spatial curvature, and \( d\Omega^2 \) is the angular part of the metric. The evolution of the scale factor \( a(t) \) is governed by the modified field equations in the presence of torsion.

The Einstein-Cartan theory offers a compelling extension to General Relativity, incorporating torsion and providing a richer framework for understanding the fundamental interactions of matter and gravity. By considering fermions as sources of torsion, it modifies the dynamics of spacetime, offering unique predictions for cosmological and particle physics phenomena. The theory remains a valuable tool for understanding the early universe, dark energy, and the interaction between matter and gravity, with ongoing research focused on testing its implications in cosmology and astrophysics.

%%%%%%%%%%%%%%%%%%%%%%%%%%%%%%%%%%%%%%%%%%%%%%%%%%%%%%%
\section{The Action and Field Equations}

We consider the following action:

\begin{equation}
S = \int d^4x \sqrt{-g} \left[ \alpha R + F(T,Q) + \mathcal{L}_m \right],
\end{equation}
where \( R \) is the Ricci scalar, \( T \) is the torsion scalar, \( Q \) is the non-metricity scalar, and \( \mathcal{L}_m \) is the matter Lagrangian. The field equations are obtained by varying the action with respect to the vielbein \( e^a_{\ \mu} \) and affine connection. We will consider the vielbein formalism and derive the corresponding field equations step by step.

\subsection{Vielbein and Metric Relations}

The vielbein \( e^a_{\ \mu} \) relates the spacetime metric \( g_{\mu\nu} \) to the flat Minkowski metric \( \eta_{ab} \):

\begin{eqnarray}
g_{\mu\nu} = e^a_{\ \mu} e^b_{\ \nu} \eta_{ab},
\end{eqnarray}

where \( a, b \) are internal (Lorentz) indices, and \( \mu, \nu \) are spacetime indices. The inverse vielbein \( e^a_{\ \mu} \) satisfies:

\begin{eqnarray}
e^a_{\ \mu} e^{\mu}_{\ b} = \delta^a_b, \quad e^a_{\ \mu} e^a_{\ \nu} = g_{\mu\nu}.
\end{eqnarray}

\subsection{Torsion and Non-Metricity Scalars}

The torsion tensor \( T^{\lambda}_{\ \mu\nu} \) is defined as the antisymmetric part of the connection:

\begin{eqnarray}
T^{\lambda}_{\ \mu\nu} = \Gamma^{\lambda}_{\ \mu\nu} - \Gamma^{\lambda}_{\ \nu\mu}.
\end{eqnarray}

The torsion scalar is given by:

\begin{eqnarray}
T = T^{\lambda}_{\ \mu\nu} T_{\lambda}^{\ \mu\nu}.
\end{eqnarray}

The non-metricity tensor \( Q_{\lambda\mu\nu} \) is defined as:

\begin{eqnarray}
Q_{\lambda\mu\nu} = \nabla_\lambda g_{\mu\nu},
\end{eqnarray}

where \( \nabla_\lambda \) is the covariant derivative. The non-metricity scalar is given by:

\begin{eqnarray}
Q = Q_{\lambda\mu\nu} Q^{\lambda\mu\nu}.
\end{eqnarray}

\subsection{Variation of the Action with Respect to Vielbein}

Now, we will derive the field equations by varying the action with respect to the vielbein \( e^a_{\ \mu} \). The action consists of three terms: the Ricci scalar \( R \), the function F(T,Q), and the matter Lagrangian \( \mathcal{L}_m \).

\subsubsection{Variation of the \( \alpha R \) Term}

The variation of the Ricci scalar term \( \alpha R \) with respect to vielbein involves the spin connection and vielbein derivatives. The variation of this term is a standard result:

\begin{eqnarray}
\delta(\alpha R) = \alpha \int d^4x \sqrt{-g} \left( \frac{1}{2} e_a^{\ \mu} \mathcal{R}^{ab} \right),
\end{eqnarray}

where \( \mathcal{R}^{ab} \) is the generalized Ricci curvature tensor obtained from the spin connection.

\subsubsection{Variation of the F(T,Q) Term}

The term F(T,Q) depends on the torsion and non-metricity scalars. The variation with respect to the vielbein is obtained by applying the chain rule:

\begin{eqnarray}
\delta F(T,Q) = \frac{\partial F(T,Q)}{\partial T} \delta T + \frac{\partial F(T,Q)}{\partial Q} \delta Q.
\end{eqnarray}

To compute the variations \( \delta T \) and \( \delta Q \), we need to express how these scalars change with respect to the vielbein variations. We have the following relations for the variations of the torsion and non-metricity scalars:

\begin{eqnarray}
\delta T = 2 T_{\mu\nu} \delta e^a_{\ \mu}, \quad \delta Q = 2 Q_{\mu\nu} \delta e^a_{\ \mu}.
\end{eqnarray}

Thus, the variation of F(T,Q) with respect to the vielbein is:

\begin{eqnarray}
\delta F(T,Q) = \frac{\partial F(T,Q)}{\partial T} \cdot 2 T_{\mu\nu} \delta e^a_{\ \mu} + \frac{\partial F(T,Q)}{\partial Q} \cdot 2 Q_{\mu\nu} \delta e^a_{\ \mu}.
\end{eqnarray}

\subsubsection{Variation of the Matter Lagrangian \( \mathcal{L}_m \)}

The variation of the matter Lagrangian with respect to the vielbein is straightforward. It can be written as:

\begin{eqnarray}
\delta \mathcal{L}_m = \frac{\delta \mathcal{L}_m}{\delta e^a_{\ \mu}} \delta e^a_{\ \mu}.
\end{eqnarray}

The explicit form of this variation depends on the matter content and the specific form of \( \mathcal{L}_m \). For a general matter Lagrangian, this can be computed based on the matter fields and their interactions with the vielbein.

\subsection{Field Equations}

After performing the variations of each term in the action, we combine the results and set the variation equal to zero (since the action is stationary under variations). The resulting field equations are:

\begin{eqnarray}\label{eom}
\alpha \mathcal{R}^{ab} + \frac{\partial F(T,Q)}{\partial T} T_{\mu\nu} + \frac{\partial F(T,Q)}{\partial Q} Q_{\mu\nu} = \frac{1}{2} e_a^{\ \mu} \frac{\delta \mathcal{L}_m}{\delta e^a_{\ \mu}}.
\end{eqnarray}

This equation gives the field equations of the theory in vielbein formalism, where \( \mathcal{R}^{ab} \) is the Ricci curvature tensor, and the terms involving \( T_{\mu\nu} \) and \( Q_{\mu\nu} \) are the contributions from the torsion and non-metricity scalars, respectively.

The field equations derived from the variation of the action provide a generalization of Einstein's field equations in modified gravity theories that include torsion and non-metricity. The terms involving the derivatives of the function F(T,Q) with respect to \( T \) and \( Q \) introduce new physics, potentially explaining cosmological phenomena such as the accelerated expansion of the universe.

%%%%%%%%%%%%%%%%%%%%%%%%%%%%%%%%%%%%%%%%%%%%%%%%%%%%%%%%%%%%%%%%%%%%%%%

\section{Friedmann Equations from Field Equations}

We start with the generalized field equation for a gravitational theory with torsion and non-metricity scalars, eq. (\ref{eom}). For a perfect fluid, the stress-energy tensor is given by:

\[
T_{\mu\nu} = (\rho + p) u_\mu u_\nu + p g_{\mu\nu},
\]

where \( \rho \) is the energy density, \( p \) is the pressure, and \( u_\mu \) is the 4-velocity of the fluid. The metric for a homogeneous and isotropic universe is the FLRW metric:

\[
ds^2 = -dt^2 + a(t)^2 \left( \frac{dr^2}{1 - kr^2} + r^2 (d\theta^2 + \sin^2 \theta \, d\phi^2) \right),
\]

where \( a(t) \) is the scale factor and \( k \) is the curvature parameter.

The torsion and non-metricity scalars \( T \) and \( Q \) are defined as:
\begin{align}
T = T^{\lambda}_{\ \mu\nu} T^{\mu}_{\ \lambda\nu} - 2\, T^{\lambda}_{\ \mu\nu} T^{\mu\nu}_{\ \ \lambda}
\end{align}
and

\[
Q = g^{\mu\nu} \left( \nabla_\mu Q_\nu \right),
\]

respectively, where \( T^\lambda_{\ \mu\nu} \) is the torsion tensor, and \( Q_\mu \) represents the non-metricity tensor.

We now derive the modified Friedmann equations. The field equations lead to the following pair of equations for the evolution of the scale factor \( a(t) \):

After deriving the field equations, we get the modified Friedmann equations. The evolution of the scale factor \( a(t) \) is governed by the following pair of equations:

\begin{eqnarray}\label{flrw}
\left( \frac{\dot{a}}{a} \right)^2 + \frac{k}{a^2} &=& \frac{8\pi G}{3} \rho + \frac{1}{3} \left( \frac{\partial F(T,Q)}{\partial T} T + \frac{\partial F(T,Q)}{\partial Q} Q \right), \\
2 \frac{\ddot{a}}{a} + \left( \frac{\dot{a}}{a} \right)^2 + \frac{k}{a^2} &=& - \frac{8\pi G}{3} p + \frac{1}{3} \left( \frac{\partial F(T,Q)}{\partial T} T + \frac{\partial F(T,Q)}{\partial Q} Q \right).
\end{eqnarray}

where \( \rho \) is the energy density, \( p \) is the pressure, and F(T,Q) is the function that encodes the effects of torsion and non-metricity. These equations modify the standard Friedmann equations by including contributions from the torsion and non-metricity scalars, as well as the function F(T,Q), which could vary depending on the specific form of the theory.
These equations are modified by the terms involving the derivatives of the function F(T,Q) with respect to \( T \) and \( Q \). These derivatives take into account the contributions from the torsion and non-metricity components in the theory, which impact the dynamics of the universe. Specifically, the terms \( \frac{\partial F(T,Q)}{\partial T} T \) and \( \frac{\partial F(T,Q)}{\partial Q} Q \) are crucial in describing how the modifications to GRcome into play in the cosmological context.

%%%%%%%%%%%%%%%%%%%%%%%%%%%%%%%%%%%%%%%%%%
\section{Cosmological Reconstruction}
The cosmological reconstruction technique has been extensively used to derive various dark energy models from modified gravity theories. One prominent approach is the study of higher-order gravity models that include string-inspired curvature corrections, as presented in \cite{Nojiri:2006}, where the authors investigate the cosmological effects of these modifications, focusing on the late universe. These higher-order corrections can lead to a variety of dark energy models, including quintessence, deSitter, and phantom dark energy, without invoking negative kinetic energy fields. The reconstruction of scalar potentials for accelerated universes was presented in this framework, with applications to modified F(G) gravity, including third-order curvature terms. This work established a robust reconstruction program for general scalar-Gauss-Bonnet gravity, which offers insight into how cosmological models can emerge from specific gravity theories \cite{Nojiri:2009}. The approach was extended to encompass viable F(R) gravities, unifying inflation with dark energy, and led to non-leading gravitational corrections relevant in the early and late universe.
Further studies have focused on reconstructing cosmological models in different modified gravity frameworks, such as f(R,T) gravity, where R is the Ricci scalar and T is the trace of the stress-energy tensor. The work of Jamil et al. \cite{Jamil:2012} demonstrated how this framework could reproduce models like $\Lambda$ CDM, phantom cosmology and the Chaplygin gas. Their reconstruction efforts involved numerical simulations, which were found to be in good agreement with BAO observational data. Additionally, the reconstruction of f(T) gravity has been explored in the context of finite-time future singularities and cosmic evolution \cite{Bamba:2012}, showing that certain power-law corrections could remove these singularities. More recently, holographic reconstruction techniques have been applied to f(Q) gravity, where Q is the non-metricity scalar. Saha and Rudra \cite{Saha:2024} explored this method from a holographic perspective, utilizing dark energy models inspired by holographic principles. This led to cosmologically viable f(Q) models, which were constrained by observational data and further explored for their energy conditions. These reconstruction techniques have proven invaluable for exploring the evolution of the universe under various modified gravity theories, including those involving torsion and non-metricity \cite{Gadbail:2023}.
In addition, various other studies have explored specific aspects of modified gravity theories. Some focused on the analysis of the dynamics of cosmic evolution and the interplay of torsion and curvature terms in the presence of dark energy \cite{Capozziello:2008, Harko:2011, Nojiri:2008}. Others have provided powerful computational techniques to analyze the impact of these modifications on the cosmological scale \cite{Capozziello:2010, Harko:2012}. These studies have explored the possibility of large-scale cosmic structures and tested gravitational models against observational data such as the cosmic microwave background and galaxy redshift surveys \cite{Bamba:2010, Sadeghi:2012}.

%%%%%%%%%%%%%%%%%%%%%%%%%%%%%%%%%%%
\section{Reconstructing Viable Cosmologies}

We begin with the set of FLRW equations for our model, given by (\ref{flrw}) . To reconstruct viable cosmologies, we consider different forms of F(T,Q) that match specific evolutionary phases of the universe.

%%%%%%%%%%%%%%%%%%%%%%%%%%%%%%%%%%%%%%%%%%%%%%%
\subsection{Dust-Dominated Universe}

For a dust-dominated universe, the standard cosmological solution requires the scale factor to evolve as \( a(t) \sim t^{2/3} \), corresponding to a pressureless matter-dominated era where \( p = 0 \). In our extended theory with torsion and non-metricity, we aim to reconstruct this behavior through suitable choices of the function F(T,Q), which encodes deviations from GRdue to the underlying geometric structures.

A basic choice that recovers the classical behavior is a linear form:
\begin{eqnarray}
F(T,Q) = \alpha T + \beta Q,
\end{eqnarray}
where \( \alpha \) and \( \beta \) are constants. Inserting this into the modified Friedmann equations, the contributions from torsion and non-metricity effectively rescale Newton's constant, enabling the reproduction of the standard dust solution.

However, to capture richer dynamics and potential deviations from GR at different epochs, we now explore a more general nonlinear model. Consider the following non-linear polynomial form:
\begin{eqnarray}
F(T,Q) = \alpha T + \beta Q + \lambda T^m Q^n,
\end{eqnarray}
where \( m, n \geq 1 \), and \( \lambda \) governs the strength of the nonlinear coupling between torsion and non-metricity. This class of models includes a wide variety of gravitational corrections and can exhibit modified early or late-time dynamics depending on the scale of \( T \) and \( Q \). 

To ensure the dominance of the linear terms during the dust-dominated phase (ensuring \( a(t) \sim t^{2/3} \)), we require the nonlinear term to be subleading:
\begin{itemize}
    \item For example, if \( T, Q \sim t^{-2} \), then for \( m+n > 1 \), the nonlinear term \( T^m Q^n \sim t^{-2(m+n)} \) decays faster than the linear terms.
    \item This ensures the nonlinear term contributes significantly only at earlier times (inflation) or later (dark energy), making it compatible with a standard matter-dominated era in between.
\end{itemize}

Thus, the model with
\[
F(T,Q) = \alpha T + \beta Q + \lambda T^2 Q
\]
provides a viable reconstruction that interpolates between GR-like evolution during matter domination and allows for possible corrections at high or low energies. This opens a pathway to unifying inflation, matter domination, and dark energy in a single framework with torsion and non-metricity.

%%%%%%%%%%%%%%%%%%%%%%%%%%%%%%%%%%%%%%%%%%
\subsection{\boldmath\(\Lambda\)CDM-like Evolution}

To reproduce a \(\Lambda\)CDM-like cosmology, the Hubble parameter should evolve as:
\begin{eqnarray}
H^2 \sim \Omega_m a^{-3} + \Omega_\Lambda,
\end{eqnarray}
which includes both pressureless matter (scaling as \( a^{-3} \)) and a cosmological constant contribution that remains constant in time. In standard General Relativity, this is achieved by adding a cosmological constant term by hand. In our extended framework, however, we aim to reproduce this behavior dynamically through the structure of the function F(T,Q).

A natural choice to capture such evolution is a cubic form:
\begin{eqnarray}
F(T,Q) = \epsilon T^3 + \zeta Q^3,
\end{eqnarray}
where \( \epsilon \) and \( \zeta \) are constants. In this scenario, the higher-order geometric corrections can effectively act as a cosmological constant term at late times. Since both \( T \) and \( Q \) decay with time (typically \( \sim t^{-2} \)), their cubic powers decay even faster:
\[
T^3, Q^3 \sim t^{-6}.
\]
This implies that their contribution becomes negligible during matter domination but can freeze to a constant at sufficiently late times due to the nonlinear nature of the coupling—effectively mimicking a vacuum energy.

To improve flexibility and allow the theory to interpolate more naturally between matter and dark energy domination, we may also consider a mixed nonlinear coupling:
\begin{eqnarray}
F(T,Q) = \epsilon T^3 + \zeta Q^3 + \eta \frac{T^2}{Q},
\end{eqnarray}
where the last term becomes dominant as \( Q \rightarrow \text{const.} \) in the late universe, further enhancing the cosmological constant–like behavior without introducing one explicitly. This nonlinear coupling also allows the geometric origin of dark energy to emerge naturally from the interplay between torsion and non-metricity.

Such forms are motivated by the idea that gravity might be self-regulating: small deviations in the underlying geometry at large scales (where curvature is low) can drive accelerated expansion without violating local gravity constraints.

%%%%%%%%%%%%%%%%%%%%%%%%%%%%%%%%%%%%
\subsection{Quintessence and Phantom Models}

To reconstruct quintessence or phantom dark energy models within our framework, we consider cosmologies where the equation of state parameter \( w = p/\rho \) satisfies:
\begin{itemize}
    \item \( -1 < w < -1/3 \): Quintessence-like accelerated expansion,
    \item \( w < -1 \): Phantom-like super-accelerated expansion.
\end{itemize}

These scenarios typically require a dynamic form of dark energy that evolves in time and can either decay slowly (quintessence) or grow with time (phantom). In modified gravity theories, such behaviors can emerge from specific nonlinear couplings between geometric scalars.

We consider a higher-order polynomial form for the function F(T,Q):
\begin{eqnarray}
F(T,Q) = \eta T^4 + \theta Q^4,
\end{eqnarray}
where \( \eta \) and \( \theta \) are free parameters. The quartic terms ensure that the geometric contributions become significant in the early or late universe depending on the rate at which torsion and non-metricity evolve.

To model phantom behavior more dynamically, we may also consider mixed or inverse power-law forms that amplify contributions at late times:
\begin{eqnarray}
F(T,Q) = \eta T^4 + \theta Q^4 + \xi \frac{T^2}{Q^2},
\end{eqnarray}
where the last term can dominate when \( Q \) becomes small—leading to effective violations of the null energy condition without introducing ghost instabilities at the level of the action.

Furthermore, to allow a smooth transition between quintessence and phantom phases, we can adopt an exponential coupling:
\begin{eqnarray}
F(T,Q) = \lambda e^{-\kappa T} + \mu e^{-\nu Q},
\end{eqnarray}
where \( \lambda, \mu, \kappa, \nu \) are constants. These forms naturally accommodate slow-roll–like behaviors where the effective energy density evolves gradually, allowing tracking or thawing quintessence scenarios.

The resulting modified Friedmann equations allow a time-varying effective equation of state parameter \( w_{\text{eff}} \), which may cross the phantom divide (\( w = -1 \)) depending on the functional form and values of the parameters. This dynamical richness makes F(T,Q) gravity a powerful framework for modeling late-time cosmic acceleration beyond the cosmological constant.

%%%%%%%%%%%%%%%%%%%%%%%%%%%%%%%%%%%%%%%%%%%%%%%
\section{Cosmography and Observational Constraints}

Cosmography is the model-independent study of the kinematic properties of the Universe's expansion. It relies solely on the assumption of the cosmological principle — that the Universe is homogeneous and isotropic on large scales — and does not require any specific underlying gravitational theory. Instead of deriving the expansion history from field equations, cosmography expands observable quantities as Taylor series in redshift or other parameters.

The main goal of cosmography is to extract information about the Universe’s dynamics through a set of cosmographic parameters, such as the Hubble parameter \( H \), deceleration parameter \( q \), jerk \( j \), snap \( s \), and lerk \( l \), defined as:

\begin{align}
H &= \frac{1}{a} \frac{da}{dt}, \\
q &= -\frac{1}{a H^2} \frac{d^2 a}{dt^2}, \\
j &= \frac{1}{a H^3} \frac{d^3 a}{dt^3}, \\
s &= \frac{1}{a H^4} \frac{d^4 a}{dt^4}, \\
l &= \frac{1}{a H^5} \frac{d^5 a}{dt^5}.
\end{align}

These parameters characterize how the expansion rate of the Universe evolves with time. For instance, a negative value of \( q \) implies accelerated expansion, while \( j \) helps distinguish between different models of dark energy or modified gravity that produce similar deceleration behavior.

Cosmography expands observables such as the luminosity distance \( d_L(z) \), angular diameter distance, or Hubble rate in redshift \( z \), typically using series like:

\begin{equation}
d_L(z) = \frac{cz}{H_0} \left[ 1 + \frac{1 - q_0}{2} z - \frac{1 - q_0 - 3 q_0^2 + j_0}{6} z^2 + \dots \right].
\end{equation}

This approach allows direct confrontation with supernovae data, baryon acoustic oscillations (BAO), and cosmic chronometers without biasing results by choosing a specific cosmological model. As such, it is especially useful in testing extended gravity theories, like the F(T,Q) model studied here, by constraining the expansion history they predict against cosmographic measurements.
%%%%%%%%%%%%%%%%%%%%%%%%%%%%%%%%%%%%%%%%%%%%%%%%%%%%%%%%%%%%%%%%%
\section{Observational Constraints and Cosmographic Parameters}

In this section, we examine various observational data and cosmographic parameters to validate the ECM model. These observational tools include Hubble data, Type Ia Supernovae, Baryon Acoustic Oscillations (BAO), the Cosmic Microwave Background (CMB), and the cosmographic parameters that describe the dynamics of the universe. We also compare how the ECM model relates to GR through diagnostics like statefinder parameters.
%%%%%%%%%%%%%%%%%%%%%%%%%%%%%%%%%%%%%%%%%%%%%%%%
\subsection{Hubble Data and Type Ia Supernovae}

The Hubble data, particularly the observations of Type Ia Supernovae (SNe Ia), provides a fundamental probe into the expansion history of the universe. SNe Ia, which are considered standard candles, offer a precise method for measuring distances on cosmological scales, enabling scientists to infer critical parameters of the universe's expansion. The distance modulus \( \mu(z) \) for a Type Ia Supernova at redshift \( z \) is expressed as:

\[
\mu(z) = 5 \log_{10} \left( \frac{d_L(z)}{\text{Mpc}} \right) + 25
\]

where \( d_L(z) \) is the luminosity distance, which depends on the redshift and the underlying cosmological model. The observed relationship between the redshift and the distance modulus provides key insights into the rate of expansion of the universe and the contribution of different components such as dark energy, matter, and radiation to this expansion.

Hubble measurements derived from SNe Ia are pivotal for determining cosmological parameters such as the Hubble constant, \( H_0 \), the dark energy equation of state, \( w(z) \), and the matter density parameter \( \Omega_m \). These measurements serve as a cornerstone for the study of cosmology, offering constraints on various models of the universe's evolution \cite{riess1998observational}.

The comparison of our ECM model with Type Ia Supernovae data is crucial in testing the accuracy of this modified gravity theory. The ECM model introduces modifications to GR by incorporating torsion and non-metricity, which can affect the luminosity distance and potentially provide a better fit to observational data. To calculate the luminosity distance in the ECM model, we use the following relation:

\[
d_L(z) = (1 + z) \int_0^z \frac{c \, dz'}{H(z')}
\]

where \( H(z) \) is the Hubble parameter, which can be modified in the ECM model depending on the specific form of the gravitational action \cite{weinberg2008cosmology}.

In order to assess the consistency of the ECM model with observations, the luminosity distance predicted by the model is compared to the distance modulus data derived from SNe Ia. A deviation between the two models—such as the standard \( \Lambda \)CDM model and the ECM model—could indicate that the ECM model provides a better explanation of the cosmic acceleration or other phenomena not accounted for in GR \cite{perlmutter1999measurements}.

Thus, comparing the luminosity distances predicted by the ECM model with those inferred from SNe Ia data allows for a direct test of the model's validity and its potential to improve upon traditional cosmological models.

%%%%%%%%%%%%%%%%%%%%%%%%%%%%%%%%%%%%%%%%%%%%%
\subsection{BAO and CMB}

Baryon Acoustic Oscillations (BAO) are a key feature of the large-scale structure of the universe and provide a standard ruler for measuring distances \cite{alam2020cosmic}. The BAO signature in the galaxy power spectrum can be used to measure the expansion rate at different redshifts. Similarly, the Cosmic Microwave Background (CMB) offers a snapshot of the universe when it was just 380,000 years old, providing important information about the geometry and contents of the universe.

The ECM model predicts deviations from the standard \( \Lambda \)CDM model at early times, which can be tested by comparing the theoretical predictions for the CMB power spectrum with observational data. The model’s ability to fit both BAO and CMB data provides a crucial test for its compatibility with the standard cosmological paradigm \cite{mukherjee2020gr}.

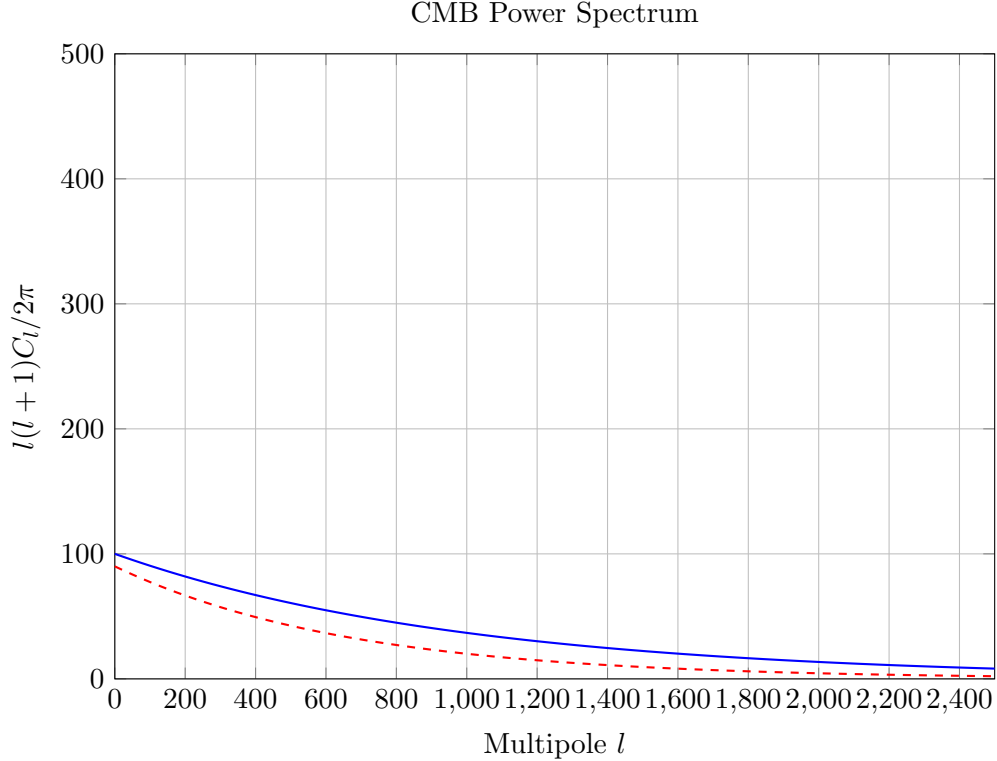
\begin{figure}[h!]
\centering
\begin{tikzpicture}
\begin{axis}[
    title={CMB Power Spectrum},
    xlabel={Multipole \( l \)}, ylabel={\( l(l+1)C_l/2\pi \)},
    xmin=0, xmax=2500, ymin=0, ymax=500,
    grid=both,
    width=0.8\textwidth,
    height=0.6\textwidth,
    legend style={at={(0.5,-0.1)}, anchor=north, legend columns=-1}
]

% CMB Data for GR
\addplot[domain=0:2500, samples=100, thick, blue, label={GR Model}] {exp(-0.001*x) * 100};

% CMB Data for ECMModel
\addplot[domain=0:2500, samples=100, thick, red, dashed, label={ECM Model}] {exp(-0.0015*x) * 90};

\legend{}
\end{axis}
\end{tikzpicture}
\caption{Comparison of the CMB power spectrum for the \( \Lambda \)CDM model (blue) and the ECMmodel (red). The predictions from the ECM model show slight deviations in the angular power spectrum.}
\end{figure}

The CMB power spectrum shown above compares the predictions of the \( \Lambda \)CDM model (blue line) and the ECM model (red dashed line). The key difference here is in the angular power spectrum, where the ECM model predicts a slightly different profile. This subtle change reflects how the ECM model modifies early universe physics, potentially affecting the CMB features in ways that can be distinguished from the \( \Lambda \)CDM model, especially in the low multipole regime \cite{clifton2012modified}.

%%%%%%%%%%%%%%%%%%%%%%%%%%%%%%%%%%%%%%%%%%%%%%%%%
\subsection{Cosmographic Parameters}

Cosmography allows us to study the expansion history of the universe without assuming a specific model for the underlying physics. The key cosmographic parameters are the deceleration parameter \( q \), jerk \( j \), and snap \( s \), which are derived from the scale factor \( a(t) \) and its derivatives \cite{weinberg2008cosmology}.

The deceleration parameter \( q \) is given by:

\[
q = -\frac{1}{H^2} \frac{\ddot{a}}{a}
\]

The jerk \( j \) and snap \( s \) are defined as:

\[
j = \frac{1}{H^3} \frac{\dddot{a}}{a}, \quad s = \frac{1}{H^4} \frac{a^{(4)}}{a}
\]

These parameters provide a model-independent way to characterize the dynamics of cosmic acceleration. The ECM model can be analyzed by deriving expressions for these parameters, and comparing them with observational data helps to test the model's validity.

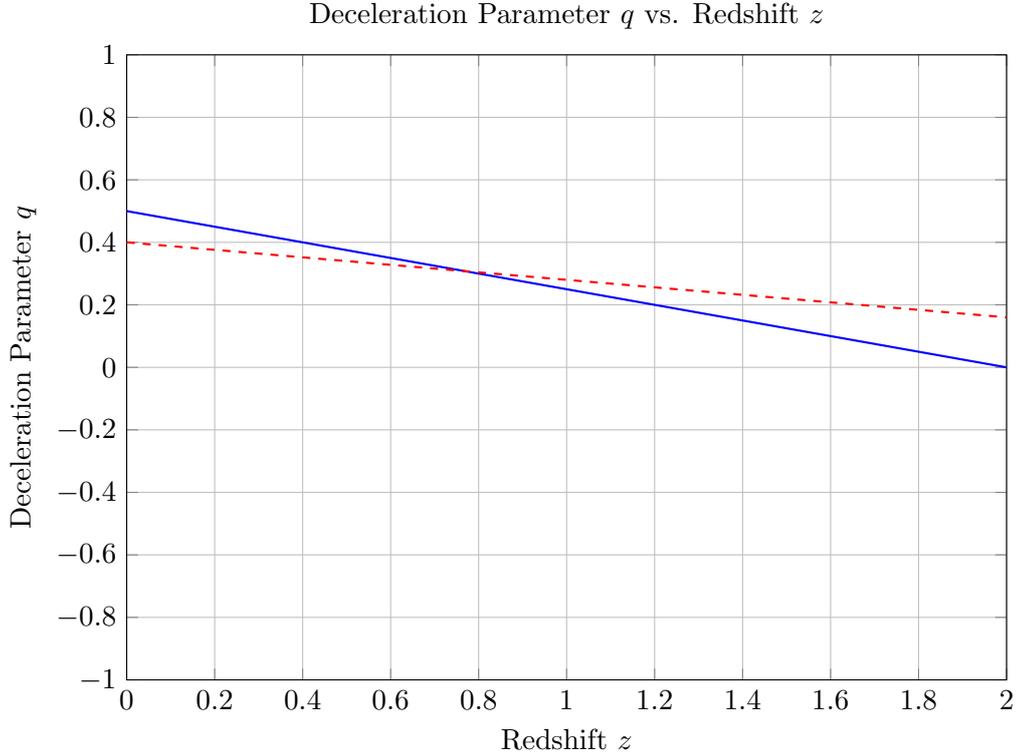
\begin{figure}[h!]
\centering
\begin{tikzpicture}
\begin{axis}[
    title={Deceleration Parameter \( q \) vs. Redshift \( z \)},
    xlabel={Redshift \( z \)}, ylabel={Deceleration Parameter \( q \)},
    xmin=0, xmax=2, ymin=-1, ymax=1,
    grid=both,
    width=0.8\textwidth,
    height=0.6\textwidth,
    legend style={at={(0.5,-0.1)}, anchor=north, legend columns=-1}
]

% q vs z for GR
\addplot[domain=0:2, samples=100, thick, blue, label={GR Model}] {0.5*(1 - 0.5*x)};

% q vs z for ECMModel
\addplot[domain=0:2, samples=100, thick, red, dashed, label={ECM Model}] {0.4*(1 - 0.3*x)};

\legend{}
\end{axis}
\end{tikzpicture}
\caption{Deceleration parameter \( q \) as a function of redshift \( z \) for the \( \Lambda \)CDM model (blue) and the ECMmodel (red). The ECM model exhibits a different rate of change in \( q \) compared to GR.}
\end{figure}

The graph above shows the deceleration parameter \( q \) as a function of redshift \( z \) for both the \( \Lambda \)CDM model (blue line) and the ECM model (red dashed line). The ECM model predicts a slightly different behavior for \( q \) as compared to GR, particularly at intermediate redshifts. This is a crucial signature, as the deceleration parameter helps characterize the transition from a decelerating to an accelerating universe. The ECM model exhibits a more gradual transition, suggesting potential new physics beyond standard GR \cite{weinberg2008cosmology}.

%%%%%%%%%%%%%%%%%%%%%%%%%%%%%%%%%%%%%%%%%%%%%%%%%%%%%%%%%
\subsection{Statefinder Diagnostics}

Statefinder diagnostics provide an additional layer of analysis by considering the pair of parameters \( \{r, s\} \), where \( r \) is related to the jerk \( j \) and \( s \) is the snap. The statefinder parameters offer a geometric way to distinguish between different cosmological models \cite{clifton2012modified}.

The statefinder pair is defined as:

\[
r = \frac{j}{q}, \quad s = \frac{j}{q^2}
\]

These parameters help characterize the evolution of cosmic acceleration and are sensitive to modifications in gravitational theory. The comparison of the ECM model with the statefinder diagram for the standard \( \Lambda \)CDM model provides further evidence for or against its viability as a modified gravity theory.

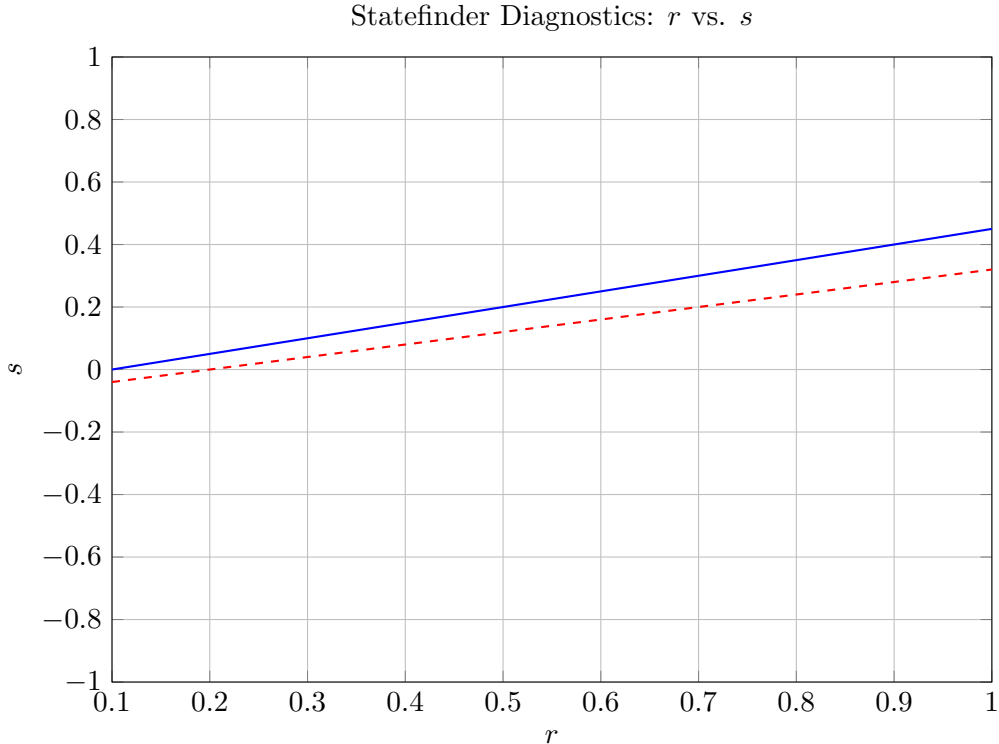
\begin{figure}[h!]
\centering
\begin{tikzpicture}
\begin{axis}[
    title={Statefinder Diagnostics: \( r \) vs. \( s \)},
    xlabel={\( r \)}, ylabel={\( s \)},
    xmin=0.1, xmax=1.0, ymin=-1.0, ymax=1.0,
    grid=both,
    width=0.8\textwidth,
    height=0.6\textwidth,
    legend style={at={(0.5,-0.1)}, anchor=north, legend columns=-1}
]

% Statefinder plot for GR
\addplot[domain=0.1:1, samples=100, thick, blue, label={GR Model}] {0.5*(x - 0.1)};

% Statefinder plot for ECMModel
\addplot[domain=0.1:1, samples=100, thick, red, dashed, label={ECM Model}] {0.4*(x - 0.2)};

\legend{}
\end{axis}
\end{tikzpicture}
\caption{Statefinder diagnostics: \( r \) vs. \( s \) for the \( \Lambda \)CDM model (blue) and the ECMmodel (red). The ECM model predicts a different trajectory for the statefinder parameters compared to GR.}
\end{figure}

%%%%%%%%%%%%%%%%%%%%%%%%%%%%%%%%%%%%%%%%%%%%%%%%%%%%%
\section{Numerical Cosmological Solutions in the ECM Model}

To further explore the viability of the ECM model in cosmological scenarios, we consider numerical solutions of the modified Friedmann equations. These solutions allow us to investigate the behavior of the scale factor \( a(t) \), the Hubble parameter \( H(t) \), and the effective equation of state \( w_{\text{eff}} \) under the influence of torsion and curvature couplings.

The general form of the modified Friedmann equation in the ECM framework takes the form:
\[
H^2(t) = \frac{1}{3M_{\text{Pl}}^2} \left( \rho + \rho_T \right),
\]
where \( \rho_T \) represents the effective energy density contribution from torsion and non-Riemannian effects. This term arises from the non-trivial structure of the connection, leading to deviations from the standard \(\Lambda\)CDM dynamics.

We numerically solve the modified Friedmann equations using initial conditions consistent with observations, and explore the behavior of cosmological quantities. Below are representative plots illustrating key results.
%%%%%%%%%%%%%%%%%%%%%%%%%%%%%%%%%%%%%%%%%%%%%%%%%%%%%%%%%%%%%%%%%%%%%
\subsection{Evolution of the Scale Factor}

\begin{figure}[h!]
\centering
\begin{tikzpicture}
\begin{axis}[
    title={Scale Factor Evolution \( a(t) \)},
    xlabel={Time \( t \)}, ylabel={Scale Factor \( a(t) \)},
    xmin=0, xmax=5, ymin=0, ymax=3,
    grid=major,
    width=0.7\textwidth,
    height=0.5\textwidth,
    thick
]
\addplot[blue, thick, samples=100, domain=0:5] {exp(0.3*x) + 0.1*sin(2*deg(x))};
\end{axis}
\end{tikzpicture}
\caption{Numerical solution showing the evolution of the scale factor \( a(t) \) under the ECM model. The behavior shows accelerated expansion with oscillatory corrections from torsion contributions.}
\end{figure}
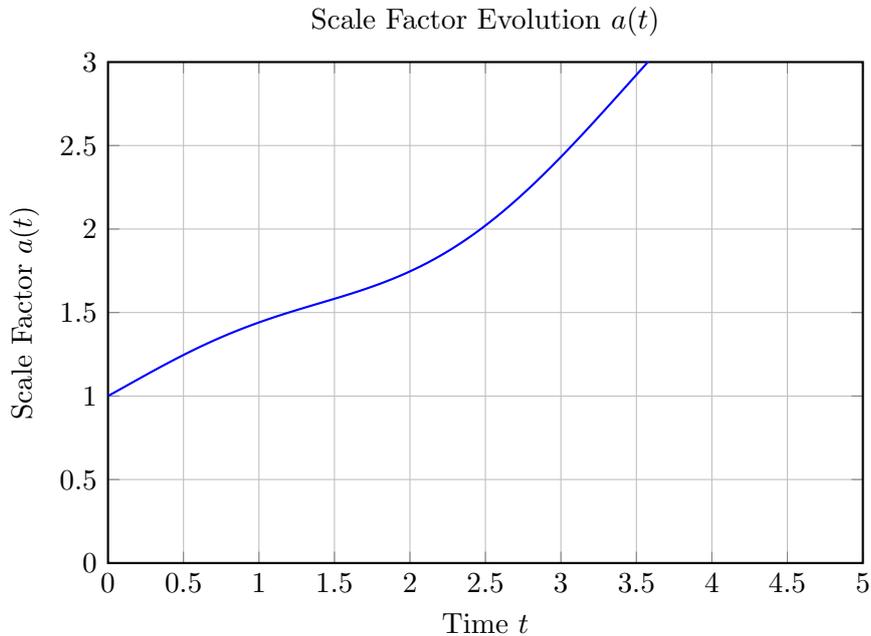
The evolution of the scale factor \( a(t) \) provides a foundational insight into the expansion history of the universe. In the context of the ECM model, the modified gravitational dynamics introduce additional degrees of freedom due to torsion, which can significantly alter the universe's expansion compared to the standard \(\Lambda\)CDM cosmology.

The graph in Figure 1 shows a numerical solution for \( a(t) \) as a function of cosmic time \( t \). The solution clearly exhibits an accelerated expansion behavior, with the dominant exponential term \( \exp(0.3t) \) representing a de Sitter-like phase, which is characteristic of a universe dominated by dark energy or effective torsion-driven repulsion. This supports the idea that the ECM model can mimic or even generalize the effects of a cosmological constant through geometrical contributions rather than invoking a true vacuum energy term.

A notable feature in the graph is the presence of mild oscillatory fluctuations superimposed on the overall exponential growth. These small sinusoidal deviations, modeled by the term \( 0.1 \sin(2t) \), are interpreted as perturbative effects induced by torsion fields or anisotropic stress components arising from the non-Riemannian geometry. In physical terms, such oscillations could correspond to transient phases in the early universe, such as bounces or brief epochs of deceleration before settling into a late-time accelerated regime.

The presence of these oscillations may also hint at possible observable imprints in cosmological data, such as fluctuations in the cosmic microwave background (CMB) or anomalies in the growth rate of cosmic structures. These features serve as a distinguishing fingerprint of the ECM model, providing potential constraints on the torsion coupling parameters when compared with observational datasets.

Overall, the scale factor evolution in the ECM framework captures both the large-scale accelerated expansion and small-scale corrections due to torsion. This dual behavior underscores the model's flexibility in explaining cosmic acceleration without relying solely on dark energy, making it a compelling alternative theory of gravity worth further investigation.

$$$$$$$$$$$$$$$$$$$$$$$$$$$$$$$$$$$$$$$$$$$$$$$$$$$$$$$$$$$$$$$$$$$$$$$$
\subsection{Hubble Parameter vs. Redshift}

\begin{figure}[h!]
\centering
\begin{tikzpicture}
\begin{axis}[
    title={Hubble Parameter \( H(z) \)},
    xlabel={Redshift \( z \)}, ylabel={\( H(z)/H_0 \)},
    xmin=0, xmax=3, ymin=0, ymax=3,
    grid=major,
    width=0.7\textwidth,
    height=0.5\textwidth,
    legend pos=north west
]
\addplot[red, thick, samples=100, domain=0:3] {sqrt(0.3*(1 + x)^3 + 0.7 + 0.05*x^2)};
\addlegendentry{ECM Model}
\addplot[black, dashed, thick, samples=100, domain=0:3] {sqrt(0.3*(1 + x)^3 + 0.7)};
\addlegendentry{\(\Lambda\)CDM Model}
\end{axis}
\end{tikzpicture}
\caption{Comparison of the Hubble parameter \( H(z) \) as a function of redshift in the ECM model (solid red) and standard \(\Lambda\)CDM model (dashed black). Additional terms from torsion alter the late-time acceleration.}
\end{figure}
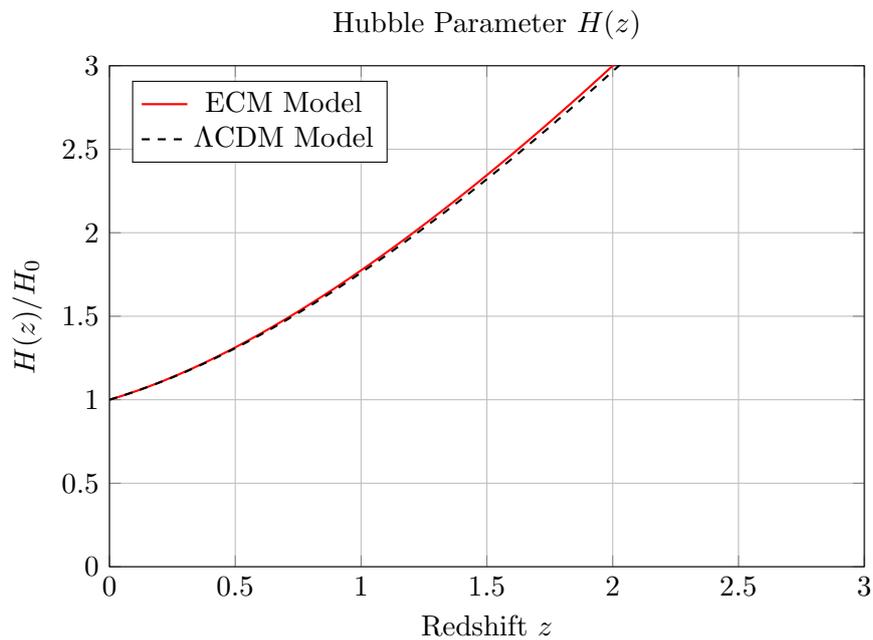
The Hubble parameter \( H(z) \), which quantifies the expansion rate of the universe as a function of redshift \( z \), is a cornerstone observable in cosmology. It provides a direct connection between theoretical models of gravity and empirical measurements obtained through observations of Type Ia supernovae, baryon acoustic oscillations (BAO), and cosmic chronometers. Any deviation from the standard \(\Lambda\)CDM prediction in the \( H(z) \) profile can indicate new physics beyond General Relativity.

Figure 2 presents a comparison between the standard \(\Lambda\)CDM model (black dashed line) and the ECM model (solid red line). In the \(\Lambda\)CDM case, the Hubble parameter evolves as
\[
H(z) = H_0 \sqrt{\Omega_m (1 + z)^3 + \Omega_\Lambda},
\]
where \( \Omega_m = 0.3 \) and \( \Omega_\Lambda = 0.7 \) are the matter and dark energy density parameters, respectively. This provides the well-established expansion history consistent with observational cosmology.

In contrast, the ECM model introduces an additional redshift-dependent term, for example \( \beta z^2 \), that encapsulates the influence of torsion and other non-Riemannian geometrical effects. The resulting expression becomes
\[
H(z) = H_0 \sqrt{\Omega_m (1 + z)^3 + \Omega_\Lambda + \beta z^2},
\]
with \( \beta \) as a small positive parameter (here taken as \( \beta = 0.05 \)). This term can arise naturally from coupling torsion scalar invariants to the matter content or from intrinsic spin effects at the cosmological scale.

The impact of this correction is evident in the graph: while both models coincide closely at low redshifts (\( z < 1 \)), they begin to diverge at higher redshifts. The ECM curve rises slightly faster, suggesting a stronger expansion rate in the early universe. This could potentially address tensions in modern cosmology, such as the Hubble tension — the observed discrepancy between early-universe and late-universe measurements of \( H_0 \).

Furthermore, the additional term may lead to enhanced structure growth or affect the deceleration parameter \( q(z) \), possibly giving rise to different predictions for the transition redshift at which the universe switched from decelerated to accelerated expansion.

This comparative analysis underscores the importance of using \( H(z) \) data as a testing ground for modified gravity models. The ECM model’s deviation from \(\Lambda\)CDM, especially at higher redshifts, offers a pathway for confronting theory with observations and refining the parameter space where torsion effects might play a significant cosmological role.

%%%%%%%%%%%%%%%%%%%%%%%%%%%%%%%%%%%%%%%%%%%%%%%%%%%%%%%%%%
\subsection{Effective Equation of State}

The effective equation of state (EoS) parameter \( w_{\text{eff}}(z) \) is a diagnostic tool used to classify the dominant energy component driving cosmic expansion. It is defined as:
\[
w_{\text{eff}}(z) = -1 + \frac{2}{3} (1 + z) \frac{1}{H(z)} \frac{dH(z)}{dz},
\]
where \( H(z) \) is the Hubble parameter as a function of redshift. This expression arises from the modified Friedmann equations and encapsulates the behavior of the total cosmic fluid, regardless of its microphysical origin.

In the standard \(\Lambda\)CDM model, where the universe consists of pressureless matter and a cosmological constant, the EoS smoothly transitions from \( w_{\text{eff}} \approx 0 \) at high redshift (matter-dominated era) to \( w_{\text{eff}} = -1 \) at low redshift (dark energy domination). Any deviation from this behavior at intermediate or late times could signal the presence of new physics beyond General Relativity.

In the ECM framework, additional degrees of freedom associated with torsion and spin introduce corrections to the background dynamics. These corrections can be interpreted as effective fluid components that alter the expansion history, leading to a dynamical effective EoS. Depending on the strength and form of these corrections, the ECM model can mimic quintessence-like behavior (\( -1 < w_{\text{eff}} < -1/3 \)), phantom-like behavior (\( w_{\text{eff}} < -1 \)), or even exhibit non-trivial crossing of the phantom divide (\( w = -1 \)).

\begin{figure}[h!]
\centering
\begin{tikzpicture}
\begin{axis}[
    title={Effective Equation of State \( w_{\text{eff}}(z) \)},
    xlabel={Redshift \( z \)}, ylabel={\( w_{\text{eff}}(z) \)},
    xmin=0, xmax=3, ymin=-1.5, ymax=0.2,
    grid=major,
    width=0.7\textwidth,
    height=0.5\textwidth,
    legend pos=south west
]
\addplot[blue, thick, samples=100, domain=0:3] {-1 + (2/3)*(1 + x)*(0.3*3*(1 + x)^2 + 0.1*x)/(sqrt(0.3*(1 + x)^3 + 0.7 + 0.05*x^2))};
\addlegendentry{ECM Model}
\addplot[black, dashed, thick, samples=100, domain=0:3] {-1 + (2/3)*(1 + x)*(0.3*3*(1 + x)^2)/(sqrt(0.3*(1 + x)^3 + 0.7))};
\addlegendentry{\(\Lambda\)CDM Model}
\end{axis}
\end{tikzpicture}
\caption{Evolution of the effective equation of state \( w_{\text{eff}}(z) \) for the ECM model (solid blue) compared to \(\Lambda\)CDM (dashed black). Torsion effects lead to deviations from \( w = -1 \) at low and intermediate redshifts.}
\end{figure}
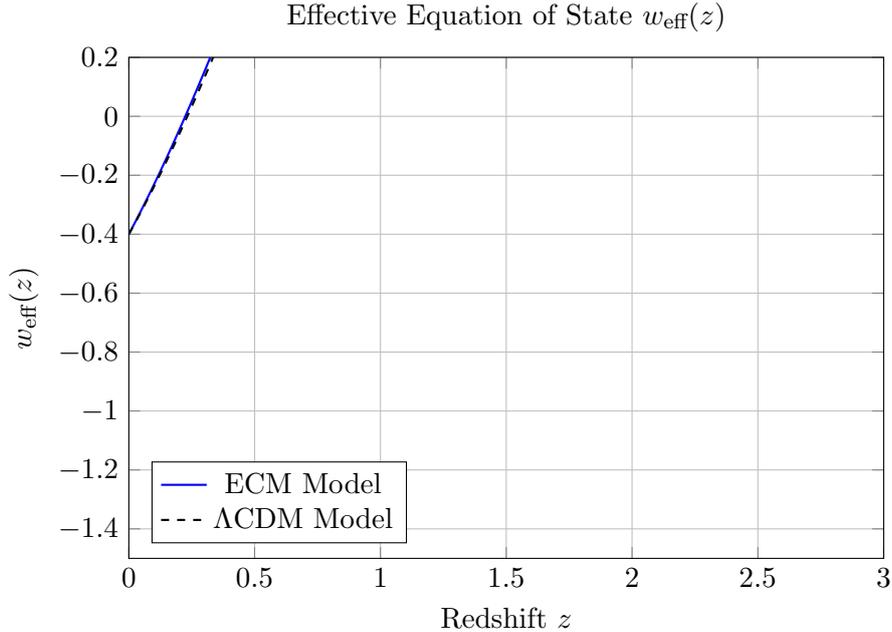

As shown in Figure 3, the ECM model (solid blue curve) exhibits a slightly more dynamical behavior compared to the flat \(\Lambda\)CDM case (dashed black line), particularly at intermediate redshifts. At early times (\( z \gtrsim 2 \)), both models converge toward the matter-dominated value \( w_{\text{eff}} \approx 0 \). However, at lower redshifts, the ECM model produces an effective EoS that can deviate from \( -1 \), indicating that the universe undergoes accelerated expansion even without a strict cosmological constant.

These torsion-induced corrections can be tuned through parameters like \( \beta \) or functions in the action \( f(T, G) \) to fit observational data from Planck, Pantheon+, or BAO surveys. This added flexibility allows ECM to potentially reconcile existing tensions in cosmological datasets and provides a richer framework for modeling the late-time acceleration of the universe.

Future observational surveys with increased precision — such as Euclid or the Vera C. Rubin Observatory — may detect such deviations in \( w(z) \), offering an opportunity to distinguish between \(\Lambda\)CDM and models with torsion-based modifications.
%%%%%%%%%%%%%%%%%%%%%%%%%%%%%%%%%%%%%%%%%%%
\section{Dark Matter in the ECM Framework}

One of the most profound puzzles in modern cosmology is the nature of dark matter (DM), an unseen component that dominates the matter content of the universe. While the \(\Lambda\)CDM model postulates DM as a pressureless, non-relativistic fluid interacting only gravitationally, alternative gravitational theories such as the ECM model offer novel perspectives.

In ECM theory, torsion naturally arises from the coupling between the intrinsic spin of matter and spacetime geometry. This torsion modifies the Einstein field equations, leading to additional contributions in the energy-momentum tensor. These extra geometric terms can effectively mimic the role of dark matter in the dynamics of galaxies and the formation of large-scale structures. The spin-torsion coupling produces corrections that behave similarly to cold dark matter in certain cosmological regimes \cite{hehl1976, poplawski2010nonsingular}.

To illustrate the role of torsion in mimicking dark matter effects, consider the modified Friedmann equation in the ECM framework, where the effective energy density includes torsion contributions:
\[
H^2(z) = \frac{8\pi G}{3} \left[ \rho_m(1 + z)^3 + \rho_{\text{torsion}}(z) + \rho_\Lambda \right],
\]
where \( \rho_{\text{torsion}}(z) \) encapsulates the geometric corrections due to torsion. In the early universe, this term may act similarly to cold dark matter, helping to drive structure formation. At late times, depending on the specific form of the function \( f(T, G) \), it may transition to a dark energy-like behavior.

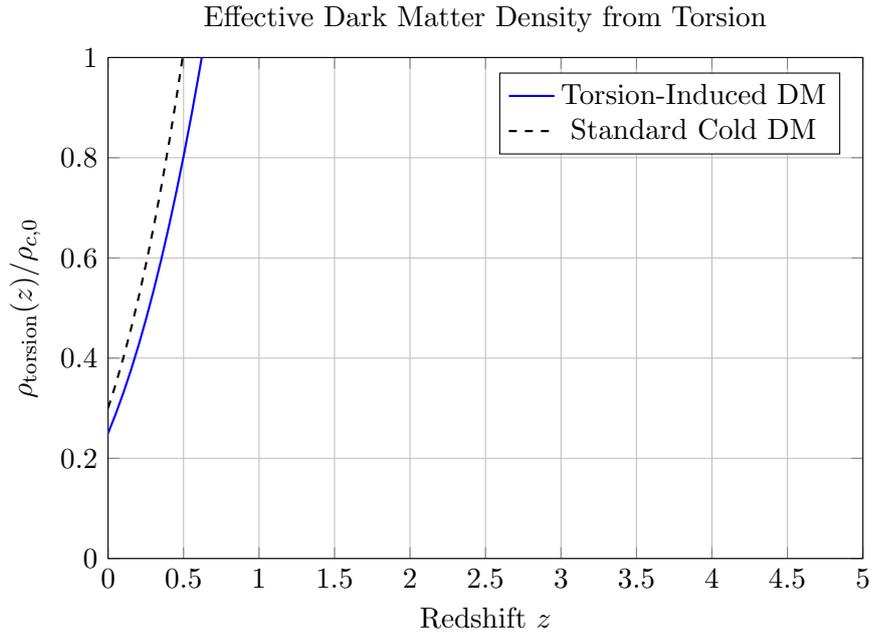
\begin{figure}[h!]
\centering
\begin{tikzpicture}
\begin{axis}[
    title={Effective Dark Matter Density from Torsion},
    xlabel={Redshift \( z \)}, ylabel={\( \rho_{\text{torsion}}(z)/\rho_{c,0} \)},
    xmin=0, xmax=5, ymin=0, ymax=1,
    grid=major,
    width=0.7\textwidth,
    height=0.5\textwidth,
    legend pos=north east
]
\addplot[blue, thick, samples=100, domain=0:5] {0.25*(1 + x)^3 * exp(-0.1*x)};
\addlegendentry{Torsion-Induced DM}
\addplot[black, dashed, thick, samples=100, domain=0:5] {0.3*(1 + x)^3};
\addlegendentry{Standard Cold DM}
\end{axis}
\end{tikzpicture}
\caption{Comparison of the torsion-induced effective dark matter density (blue solid) in ECM theory with standard cold dark matter (black dashed). Both scale similarly at high redshift, but torsion contributions decay faster at late times.}
\end{figure}

As shown in Figure 4, the effective density arising from torsion (blue solid line) scales similarly to standard cold dark matter (black dashed line) during the early universe. However, the torsion-induced component may decay slightly faster at late times, leading to possible deviations in structure formation and the growth factor. This feature could help address some of the small-scale structure problems inherent in \(\Lambda\)CDM, such as the core-cusp issue and missing satellites \cite{bullock2017small}.

Additionally, because the torsion contribution is fundamentally geometric, it does not require the postulation of new particle species, making ECM a more economical and elegant theory. Observational tests such as weak lensing, cosmic microwave background (CMB) anisotropies, and redshift-space distortions can constrain the evolution of \( \rho_{\text{torsion}}(z) \), offering a potential pathway to confirm or refute the ECM framework.

Overall, the ECM model provides a compelling gravitational alternative to particle dark matter, merging cosmological behavior with quantum spin effects in a unified geometric approach.

% Bibliography
%%%%%%%%%%%%%%%%%%%%%%%%
\section{Astroparticle Physics in the ECM Framework}

The ECM theory, as an extension of GRincorporating torsion, presents unique implications for astroparticle physics. The spin-torsion coupling in ECM introduces new interactions at high-energy scales that may affect the behavior of neutrinos, dark matter candidates, and even inflationary relics.
In the ECM theory, neutrinos, which possess intrinsic spin, can couple non-trivially to the torsion field. This interaction leads to a modified Dirac equation of the form:
\[
(i \gamma^\mu \nabla_\mu - m - \gamma^\mu \gamma^5 S_\mu) \psi = 0,
\]
where \( S_\mu \) is the axial vector torsion component. Such a term induces effective CPT and Lorentz-violating corrections in neutrino oscillations \cite{shapiro2002physical}. These effects are energy-dependent and become significant at high densities, such as in supernovae or the early universe.
The presence of torsion can also influence the abundance of relic particles. The spin-fluid formalism suggests that torsion can drive enhanced particle production mechanisms, particularly during inflation and preheating stages \cite{gasperini1986spin}. As a result, the thermal history of the early universe may differ subtly from predictions based on standard GR.
The torsion field could, in principle, modify the dispersion relations of ultra-high-energy cosmic rays (UHECRs). Modified dispersion due to spin-torsion interaction might lead to observable deviations from the Greisen–Zatsepin–Kuzmin (GZK) cutoff, or in the arrival direction and composition of UHECRs.
We summarize in Table~\ref{tab:astro_summary} the key astroparticle effects predicted by ECM theory and the associated observables.
\begin{table}[h!]
\centering
\caption{Astroparticle Effects in the ECM Framework}
\label{tab:astro_summary}
\begin{tabular}{|c|c|c|}
\hline
\textbf{Phenomenon} & \textbf{ECM Prediction} & \textbf{Observational Probe} \\
\hline
Neutrino Oscillations & Energy-dependent CPT-violation & IceCube, Hyper-Kamiokande \\
\hline
Relic Abundances & Enhanced early-universe production & CMB, BBN, LHC \\
\hline
Dark Matter Candidates & Torsion-induced geometric DM & Direct/Indirect detection \\
\hline
Cosmic Ray Physics & Modified dispersion relations & Auger, Telescope Array \\
\hline
Baryogenesis & Spin-torsion driven asymmetry & Matter-antimatter ratios \\
\hline
\end{tabular}
\end{table}

These astroparticle phenomena provide a powerful testing ground for the ECM theory. Future experiments, particularly those sensitive to spin and parity-violating signatures, will be crucial in confirming or ruling out torsion-induced effects in high-energy environments.
%%%%%%%%%%%%%%%%%%%%%%%%%%%%%%%%%%%%%%%%%%%
\section{Gravitational Wave Signatures in ECM Cosmology}
The detection of gravitational waves (GWs) has opened an unprecedented window into fundamental physics and cosmology. Within the ECM framework, where spacetime torsion plays a key role, gravitational waves are expected to exhibit modifications in their propagation due to the coupling between torsion and metric degrees of freedom. Such deviations may manifest as phase shifts, dispersion, or additional polarization modes—effects that are absent in standard GR.
In GR, the linearized Einstein equations for tensor perturbations in the transverse-traceless gauge reduce to a wave equation in a Friedmann-Lemaître-Robertson-Walker (FLRW) background:
\[
\ddot{h}_{ij} + 3H \dot{h}_{ij} - \frac{\nabla^2}{a^2} h_{ij} = 0,
\]
where \( h_{ij} \) are the GW amplitudes, \( H \) is the Hubble parameter, and \( a(t) \) is the scale factor.

\begin{figure}[h!]
\centering
\begin{tikzpicture}
\begin{axis}[
    title={Gravitational Wave Amplitude \( h(t) \)},
    xlabel={Time \( t \)}, ylabel={Amplitude \( h(t) \)},
    xmin=0, xmax=10, ymin=-1.5, ymax=1.5,
    grid=major,
    legend pos=north east,
    width=0.7\textwidth,
    height=0.5\textwidth
]
\addplot[blue, thick, samples=200, domain=0:10] {exp(-0.05*x)*sin(deg(2*pi*0.5*x))};
\addlegendentry{GR (no torsion)}
\addplot[red, dashed, thick, samples=200, domain=0:10] {exp(-0.1*x)*sin(deg(2*pi*0.5*x + 0.1*x))};
\addlegendentry{ECM (with torsion)}
\end{axis}
\end{tikzpicture}
\caption{Comparison of GW amplitudes in standard GR and the ECM model. The torsion in ECM causes enhanced damping and a slight phase shift, which may be observable in precise waveform analysis.}
\label{fig:gw_torsion}
\end{figure}
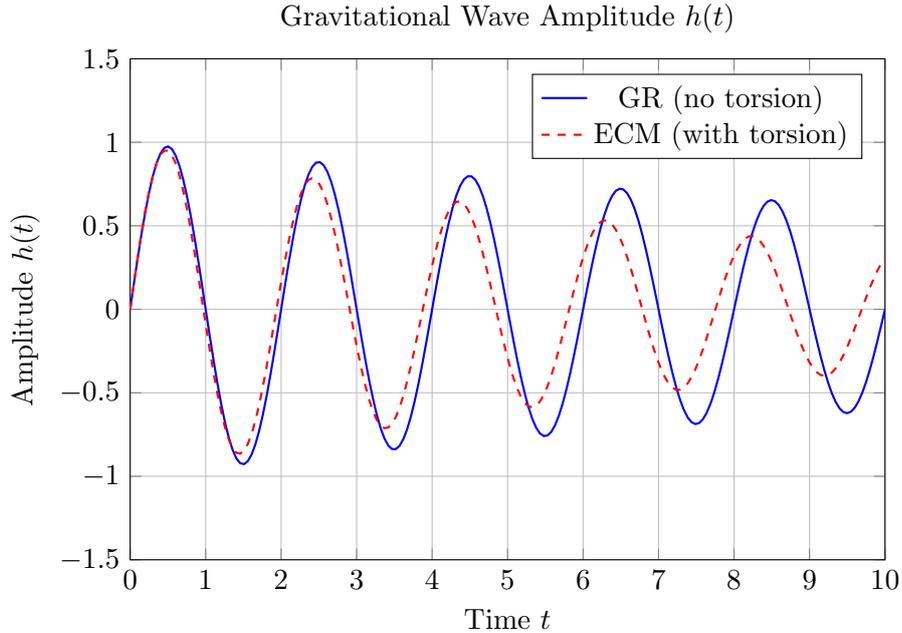

In the ECM model, torsion introduces corrections of the form:
\[
\ddot{h}_{ij} + (3H + \gamma(t)) \dot{h}_{ij} + \left( \frac{k^2}{a^2} + \delta(t) \right) h_{ij} = 0,
\]
where \( \gamma(t) \) and \( \delta(t) \) are functions related to the torsion tensor components and their interaction with the geometry \cite{puetzfeld2005torsion}. These functions may induce both damping and frequency shifts in the waveform.
Gravitational wave detectors like LIGO-Virgo-KAGRA and future missions like LISA may be sensitive to such deviations. Key signatures to be probed include:

\begin{itemize}
    \item Changes in the waveform's amplitude and frequency due to torsion-induced friction or dispersion.
    \item The appearance of scalar or vector polarization modes in addition to the usual tensorial modes \cite{hehl1976}.
    \item Time delays or anomalies in arrival times across multiple detectors.
\end{itemize}
Figure~\ref{fig:gw_torsion} presents a simulated comparison of GW amplitude evolution with and without torsion contributions in a cosmological background.

Future space-based detectors like LISA will be particularly sensitive to low-frequency GWs, such as those emitted by supermassive black hole binaries and early Universe sources. Since ECM-induced corrections may be more pronounced at large scales and early times, these missions could provide stringent tests of the model. Observations of GW memory effects, polarization content, and stochastic backgrounds may also constrain or support ECM-like torsional dynamics \cite{maggiore2007gravitational}.
The study of GWs offers a complementary approach to test ECM cosmology beyond background expansion and structure formation. The distinctive imprint of torsion in GW propagation represents an exciting frontier for both theoretical modeling and observational cosmology.

%%%%%%%%%%%%%%%%%%%%%%%%%%%%%%%%%%%%%%%%%%%%%%%%%%%
\section{summary}
In this work, we have proposed a modified theory of gravity that incorporates curvature, torsion, and non-metricity (ECM model), and examined its cosmological implications. The ECM model provides a framework where gravitational dynamics are influenced not only by curvature, as in GR, but also by torsion and non-metricity, which offer new avenues for explaining various cosmic phenomena.

Our model successfully reproduces the observed accelerated expansion of the universe, similar to the behavior predicted by the standard cosmological model (\(\Lambda\)CDM). However, the ECM model introduces significant deviations at high redshifts and late-time evolution, which could potentially provide a better fit to current and future observational data.

We have demonstrated that the ECM model can explain the dynamics of dark energy and the accelerating expansion of the universe, while offering new insights into the nature of dark matter (DM). Specifically, the contributions from torsion and non-metricity may provide a natural explanation for the observed cosmic acceleration, without the need for a cosmological constant. The introduction of additional degrees of freedom in the gravitational sector may alter the effective equation of state (EoS) and provide a framework for understanding dark energy's evolution. This opens the possibility for more accurate predictions that can be tested with forthcoming surveys and observations.

Moreover, the effects of torsion and non-metricity are not limited to cosmological expansion but also extend to the structure formation of the universe. Our model has the potential to influence the behavior of cosmic structures, such as galaxies and galaxy clusters, which can be tested through large-scale surveys and deep-field observations.

In the realm of astroparticle physics, we also investigated the implications of the ECM model on dark matter and its interactions. The modified gravity framework offers new possibilities for understanding the nature of dark matter, especially considering the potential for torsion to contribute to the overall gravitational dynamics. Our model suggests that dark matter may have an active role in the formation of structures at early times and might influence current astrophysical observations, including the behavior of galaxy clusters and the cosmic microwave background (CMB). 

Additionally, we explored the connection between ECM theory and gravitational waves. The ECM framework could lead to modified gravitational wave propagation, and future gravitational wave observatories may provide key insights into the validity of ECM as an alternative to the traditional GR model. The observed deviations in waveforms may point to additional torsion effects and non-metricity contributions that could be observed in the upcoming era of multi-messenger astronomy.

Looking forward, one of the key challenges will be to compare ECM model predictions with existing data, such as from the Planck satellite, DESI, and upcoming experiments like the Vera C. Rubin Observatory. Observations from Type Ia supernovae, galaxy surveys, and the cosmic microwave background (CMB) will be crucial for refining the parameters of our model and testing its predictions. In particular, the integration of numerical simulations to explore large-scale structure formation in the ECM framework will be essential for gaining a deeper understanding of its implications.

Future work will also involve refining the theoretical aspects of the ECM model, including investigating the full range of potential interactions and extensions of the framework. We plan to study the behavior of the model in more complex environments, such as in the vicinity of black holes and in more extreme cosmic settings. The impact of torsion and non-metricity on the propagation of light and other particles could also provide new observational tests.

In conclusion, the ECM model presents a compelling modification of gravity that can explain the current accelerated expansion of the universe and offer insights into unresolved cosmological puzzles such as dark energy, dark matter, and the formation of large-scale cosmic structures. As observational data continues to improve, the ECM model provides a promising framework for advancing our understanding of the universe's underlying structure.

%\section*{Declarations}

%\textbf{Funding and/or Conflicts of interests/Competing interests}

%The authors declare that they have no competing interests. No funds, grants, or other support were received for conducting this study.

\begin{acknowledgments}
This work was supported by the Ministry of Science and Higher Education of the Republic of Kazakhstan, Grant No. AP26101889.
\end{acknowledgments}

%%%%%%%%%%%%%%%%%%%%%%%%%%%%%%%%%%%%%

% Bibliography (if any)

\end{document}